\documentstyle[a4wide,epsf,epsfig,11pt,titlepage]{article}

\newcounter{nref}
\newcommand{\bbib}{%
  \renewcommand{\refname}{\large\bf References}%
  \begin{thebibliography}{99}%
  \setcounter{enumiv}{\thenref}}
\newcommand{\ebib}{%
  \end{thebibliography}%
  \setcounter{nref}{\arabic{enumiv}}}
\newcommand{\head}[3]{%
  \setcounter{nref}{0}%
  \thispagestyle{empty}%
  \section*{\LARGE\bf #1}
  \addcontentsline{toc}{section}{#1}%
  \large\itshape%
  #2\\\vspace{0.1pt}\\%
  #3%
  \normalsize\upshape%
  \bigskip}

  \def\vdrn{\vec{v}_{\rm dr}^{\hspace{-0.9ex}^{\circ}}}
    \def\xx{\vec{x}}
    \def\Vl{{V_{\ell}}}
\begin{document}

\sloppy

\head{The need for small-scale turbulence\\ in atmospheres of substellar objects}
     {Christiane Helling\\
      Scientific and Research Support Department,\\ ESTEC/ESA, Noordwijk, The Netherlands}

\subsection*{Abstract} Brown dwarfs and giant gas planets are substellar objects whose spectral appearance is determined by the
chemical composition of the gas and the solids/liquids in the atmosphere.  Atmospheres of substellar objects possess two major
scale regimes: large-scale convective motions + gravitational settling and small-scale turbulence + dust formation. Turbulence
initiates dust formation spot-like on small scale, while the dust feeds back into the turbulent fluid field by its strong
radiative cooling. Small, imploding dust containing areas result which eventually become isothermal. Multi-dimensional simulations
show that these small-scale dust structures gather into large-scale structures, suggesting the formation of clouds made of dirty
dust grains.  The chemical composition of the grains, and thereby the chemical evolution of the gas phase, is a function of
temperature and depends on the grain's history.

\vspace*{-14cm}
\vspace*{-2mm}
\begin{center}
\begin{small}
   {\sl Contributed talk at the\\ Workshop on Interdisciplinary Aspects of Turbulence,\\
   April 18 - 22, 2005, Castle Ringberg, Germany\\[0.1cm]

to appear in the {\sc MPA publication series} (eds. F. Kupka, W. Hillebrandt)
}
\end{small}
\end{center}
\vspace*{12cm}

\section{Introduction}

The first brown dwarf Gliese 229B has been discovered 10 years ago by direct imaging (Kulkarni \& Golimovsky 1995). These faint
($L_*=10^{-7} \ldots 10^{-1}L_{\odot}$), cool ($T_{\rm eff} = 500 \ldots 3200\,$K), and small ($M_*=0.01 \ldots 0.08 M_{\odot}$)
objects bridge the physical and chemical gap between the classical understanding of stars ($M_* > 0.08 M_{\odot}$) and planets
($M_* < 0.01 M_{\odot}$). Much closer by, direct images revels spotty, cloudy, and vortex surface pattern in the giant planet
atmospheres in our own solar system (by the Cassini and Galileo spacecrafts fly-bys of Saturn and Jupiter, respectively) which
guide our imagination for substellar but extra-solar atmospheres.  The other major source of information is the measurement of the
energy distribution of the stellar radiative flux emerging from the object's atmosphere (e.g. for Gliese 229B Oppenheimer et
al. 1998). The interpretation of the resulting spectral energy distribution demands a certain compleatness\footnote{e.g.  in
modeling the molecular regime, chemistry and hydrodynamics} of the adopted substellar atmosphere model. Substellar atmospheres,
i.e. giant gas planets and brown dwarfs, are very cool and therefore exhibit a rich molecular-- and solid--/liquid--phase
chemistry. Transitions between the phase regimes are to be expected. Therefore, models of substellar atmospheres -- as the
interface to the physical and the chemical state of the object -- need to represent the {\it circuit of dust} (Helling 1999,
Woitke \& Helling 2003) which includes the formation of dust, the chemical composition of dust and gas, gravitational settling
(rain), its feedback on the dust formation process, and element replenishment by upward convective motions in addition to
hydrodynamics and radiative transfer.  In contrast to terrestrial planets which possess their solid surface as continues source of
seed particles\footnote{Seed particles on Earth are called {\it aerosols} which are for instance volcanic dust and tire
particles, or they come from fire in the tropics and from smoke-tracks.}, the actual formation of the first (solid or liquid) surface out of
the gas phase has to be considered in substellar atmospheres. Convection is an efficient mechanism to continuously and
intermittently dredge up fresh, uncondensed gaseous material from the very bottom of the atmosphere.  The convection furthermore
serves as turbulence engine inside the atmosphere. Consequently, modeling and understanding a substellar atmosphere means to model
and to understand a reactive, dust forming, turbulent fluid field.

\begin{table}
\begin{center}
\begin{tabular}{lcl}
\hline
$\tau_{\rm sink} = \frac{H_{\rm p}}{\vdrn}$ & 15\,min $\,\ldots$ \,8\, month & dust settling\\
& {\small ($a=100\mu$m$\, \ldots \, 0.1\mu$m)}\\[0.2cm]
$\tau_{\rm conv} = \frac{l_{\rm conv}}{v_{\rm conv}}$ & 20\,min\,$\ldots$\, 3.5\,h
& large scale\\
& &  convection\\[0.2cm]
\hline
$\tau_{\rm gr} = \frac{\langle a\rangle}{\chi^{\rm net}}$ & 0.1\,s$\,\ldots\,$ 1\, 1/2\, min & dust growth\\
& {\small ($a=0.1\mu$m$\, \ldots \, 100\mu$m)}\\[0.2cm]
$\tau_{\rm wave} = \frac{L}{|U| + c_{\rm s}}$ & 0.3\,$\ldots$\,3\,s & wave propagation\\[0.2cm]

$\tau_{\rm nuc} = \frac{\rho L_0}{J_*}$ & $\approx 10^{-3}$\,s & seed formation \\[0.2cm] 
\hline \hline
$\tau_{\rm num} \approx
\frac{Re_{\rm L}^{3}}{10^5}$ & $5\cdot 10^5$\,yr & $10^3$ floating-point \\ 
&\small \hspace*{-0.3cm}($\approx 2\times$age of
mankind) & operations\\ & & per cell and $\Delta t$\\ & & with 1 gigaflop\\ 
\hline
\end{tabular} 
\caption{Time scales of the
processes involved in the {\it circuit of dust} in a substellar atmosphere.\newline Temperature $T$, density $\rho$, pressure scale
height $H_{\rm p}$, convective velocity $v_{\rm conv}$, and velocity of sound $c_{\rm s}$ are adopted from the model results by
Allard et al. (2001) and Tsuji (2002). For more details on $\tau_{\rm sink}$ see (Woitke~\& Helling 2003).}
\end{center} 
\label{tab:ts}
\end{table}

\section{Catching the small scales}

Classical models for substellar atmospheres represent the whole turbulent scale spectrum by only one scale, the mixing
length. These models have given very reasonable fits to observed spectra in certain wavelength regimes but are challenged by the
progress in observational techniques which lead to observations e.g. with higher resolutions and at longer wavelength
($\lambda>12\mu$m). Other models like Reynolds stress and LES are in progress, all of them being challenged by the closure
problem, i.e. the treatment of the smallest, unresolved scales.  

In order to provide insight and understanding of the small scale regimes of a substellar atmosphere, the interaction of turbulence
and dust formation has been studied by utilizing 1D and 2D simulations in the present work. The general phenomenology of a
substellar atmosphere model can be demonstrated by estimating the time scale of the individual processes (Table~\ref{tab:ts}). The
gravitational settling time scale of grains $\tau_{\rm sink}$ is the largest and is comparable to a typical convective mixing time
for large grains. The smallest time is needed by the formation of seed particles out of the gas phase. The dust growth time scale
is of the order of the crossing time of an acoustic wave.  Hence, two time regimes appear: (i) a quasi-static regime governed by
gravitational settling and large-scale convective motions, and (ii) a dynamic regime governed by the dust formation and
small-scale waves (turbulence). Actually, a third regime (iii) is to be faced which concerns the computing time needed to resolve
the turbulent problem (last entry Table~\ref{tab:ts}). In order to tackle regime (iii), the small-scale regime (ii) was
investigated. Here, gravitational settling can be neglected and convection acts only indirectly as turbulence driver.

\begin{figure*}
\begin{center}
\label{fig:1D}
{\epsfig{file=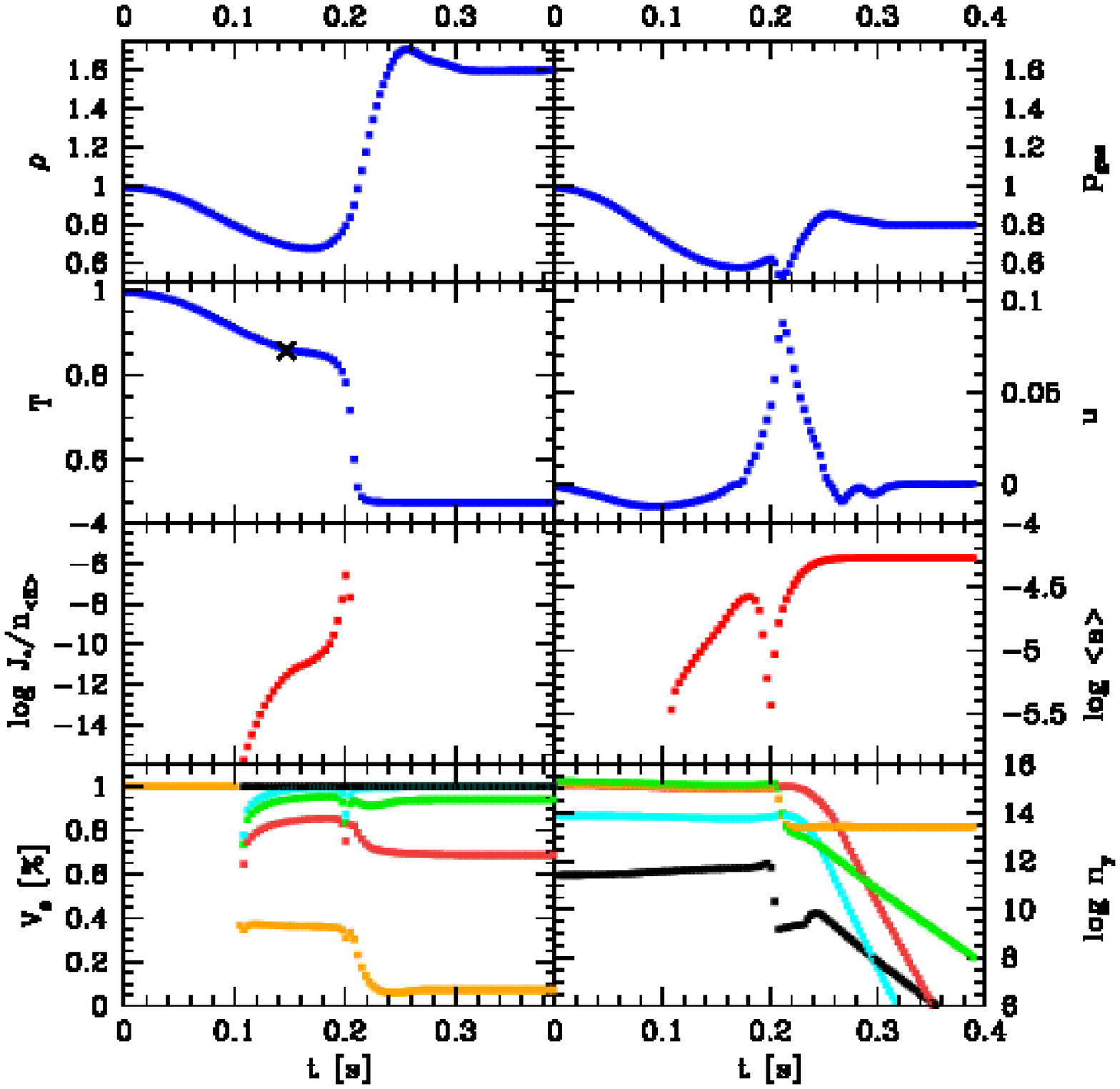, scale=0.7}}
\caption{The feedback between turbulence and dust formation in a 1D simulation of interacting turbulence elements. The time evolution of all quantities is
depicted at the site of maximum superposition of the turbulence elements.\newline 
{\bf Parameter:}
$T_{\rm ref}=1900$K, $T_{\rm RE}=1634$K, $\rho_{\rm
ref}=10^{-4}$g/cm$^3$, M=0.1 ($u_{\rm ref}=3\cdot 10^4$cm/s), $t_{\rm
ref}=3$s, $l_{\rm ref}=10^5$cm.
[numerical parameter: $N_x=500$, $N_k=500$, $\Delta x = 3.94\cdot
10^2$cm, $l_{\rm max}= l_{\rm ref}/2$]
\newline {\bf 1st row:} l.h.s. -- $\rho$ density, r.h.s. --
$p$ pressure; {\bf 2nd row:} l.h.s. -- $T$ temperature (cross -- time
of maximum wave superposition), r.h.s. -- $u$ fluid velocity; {\bf 3rd
row:} l.h.s. -- $\log J_*/n_{\rm <H>}$ nucleation rate [1/s],
r.h.s. -- $\log \langle a\rangle$ mean grain size [cm]; {\bf 4th row:}
l.h.s. -- $V_{\rm tot}=\sum V_{\rm s}$ cumulative volumes [\%] (orange
- $V_{\rm MgSiO_{3{\rm [s]}}}$, brown - $V_{\rm MgSiO_{3{\rm [s]}}} + V_{\rm
SiO_{2{\rm [s]}}}$, green - $V_{\rm MgSiO_{3{\rm [s]}}} + V_{\rm SiO_{2{\rm [s]}}} + V_{\rm
Fe_{\rm [s]}}$, light blue - $V_{\rm MgSiO_{3{\rm [s]}}} + V_{\rm SiO_{2{\rm [s]}}} +
V_{\rm Fe_{\rm [s]}} + V_{\rm Al_2O_{3{\rm [s]}}}$, dark blue - $V_{\rm
MgSiO_{3{\rm [s]}}} + V_{\rm SiO_{2{\rm [s]}}} + V_{\rm Fe_{\rm [s]}} + V_{\rm
Al_2O_{3{\rm [s]}}} + V_{\rm TiO_{2{\rm [s]}}}$), r.h.s. -- $\log n_{\rm y}$ number
density of gaseous key species for dust formation [1/cm$^3$] (orange -
Mg, brown - SiO, green - Fe, light blue - AlOH, black - TiO$_2$).}
\end{center}
\end{figure*}

The following system of dimensionless equations has been solved where Eqs.~\ref{dens}--~\ref{energy} are the equation of
continuity, of motion, and the energy equation, respectively. The source term in Eqs.~\ref{energy} is due to radiative cooling
modeled by a relaxation ansatz.  \begin{eqnarray} \label{dens} (\rho)_t + \nabla \cdot (\rho {\bf v})&=& 0\\ \label{motion} (\rho
{\bf v})_t + \nabla\cdot (\rho {\bf v}\circ {\bf v}) &=& - {\frac{1}{\bf \rm M^2}}\nabla P - \gamma {\frac{\bf \rm M^2}{\bf\rm
Fr}}\rho {\bf g}\\ \label{energy} (\rho e)_t + \nabla \cdot ({\bf v}[\rho e + P]) &=& {\bf\rm Rd}\,\kappa\, (T^4_{\rm
RE}-T^4)\\[7mm] \label{moments} (\rho L_{\rm j})_{t} + \nabla \cdot ({\bf v}\,\rho L_{\rm j}) &=& {\bf\rm Da^{\rm nuc}_{\rm d} \,
Se_{\rm j}} J_* + {\bf\rm Da^{\rm gr}_{\rm d}}\, \displaystyle{ \frac{\rm j\chi^{\rm net}}{3}} \rho L_{{\rm j}-1}\\[-2mm]
\label{consum} (\rho \epsilon_{\rm x})_t + \nabla \cdot ({\bf v}\,\rho \epsilon_{\rm x}) &=&\!\!\!\!-\!\sum_{\rm r=1}^{\rm R}
(\nu^{\rm nuc}_{\rm r}\,{\bf\rm El\,Da^{\rm nuc}_{\rm d}} \sqrt[3]{36\pi}N_{\rm l} \; J_* + \,\nu^{\rm gr}_{\rm r}\,{\bf\rm
El\,Da^{\rm gr}_{\rm d}}\;n_{\rm x, r} v_{\rm rel, x} \alpha_{\rm r} \, \rho L_2 ) \end{eqnarray} Equations~\ref{moments} ($j=0,
1, 2, 3$; $\rho L_j(\xx,t) = \int_{\rm \Vl}^{\infty} f(V,\xx,t)\,V^{\rm j/3}\,dV$ dust moments, $f(V,\xx,t)$ grain size
distribution function) model the dust formation as two step-process, namely, seed formation and mantle growth/evaporation being
the first and the second source term, respectively. Equations~\ref{consum} are element conservation equation for each chemical
element $\epsilon_{\rm x}$ (x= Mg, Si, O, Fe, Al, Ti) involved in the dust formation processes, hence each source term in
Eqs.~\ref{moments} is a sink for Eqs.~\ref{consum} (Helling et al. 2001 for details). A strong coupling exists between
Eqs.~\ref{dens}--~\ref{energy} (5 equations) and Eqs.~\ref{moments},~\ref{consum} (11 equations) due to the dust opacity $\kappa$ since it changes by order of
magnitudes if dust forms.

\subsection{Turbulence $\leftrightarrow$ Dust formation}\label{sec:td}

The feedback between turbulence and dust formation can be studied in detail by 1D simulations.  Interacting turbulence elements
are modeled as superimposing expansion waves.  Figure~\ref{fig:1D} demonstrates the time evolution of the
system at the site of constructive wave interaction.

\noindent \underline{Turbulence $\longrightarrow$ Dust formation:}\\ At about the time of superposition (black cross on T--curve,
l.h.s. 2nd row), the nucleation threshold temperature (here for TiO$_2$ seed formation) is crossed and dust nucleation is
initiated, hence the nucleation rate $J_*$ increases. Many solid compounds are already thermally stable at such low temperatures
which results in a very rapid growth of a mantle on the surface of the seed particles (here Mg$_2$SiO$_4$$_{\rm [s]}$,
SiO$_2$$_{\rm [s]}$, Fe$_{\rm [s]}$, Al$_2$O$_3$$_{\rm [s]}$, TiO$_2$$_{\rm [s]}$). As the amount of dust formed increases, the
opacity $\kappa$ of the dust-gas mixture increases by order of magnitudes. Therefore, the radiative cooling causes the temperature
$T$ to drop considerably (l.h.s., 2nd row). A classical instability establishes where the reason supports the cause. The cooler
the gas, the more dust forms, the faster the temperature drops etc. This run-away effect stops if all condensible material was
consumed or if the temperature is too low for further efficient nucleation. The time of maximum nucleation rate corresponds to a
minimum in mean grain size because suddenly the available gaseous material is needed for a much larger number of grains.

\noindent \underline{Dust formation $\longrightarrow$ Turbulence:}\\ The strong temperature gradient causes a strong raise in
density by a moderate pressure gradient. Without such a strong cooling, pressure equilibrium should adjust. Hence, the dust
forming areas implode and cause a considerable disturbance of the velocity field (here up to 10\%).  A feedback-loop {\it
turbulence $\Rightarrow$ dust formation $\Rightarrow$ turbulence} established as result of non-linear coupling of the model
equations.

\noindent \underline{Chemistry:}\\ Figure~\ref{fig:1D} depicts in the lowest two panels the strong feedback of the turbulent dust
formation process on the chemical composition of the dust grains and on the remaining gas phase. The cumulative dust volumes
(l.h.s., 4th row) show that the dust composition changes according to the temperature from Mg$_2$SiO$_4$$_{\rm [s]}$/SiO$_2$$_{\rm
[s]}$/Fe$_{\rm [s]}$--rich (40\%/40\%/10\%) with Al and Ti -- oxide impurities to a mean grain composition of 55\% SiO$_2$$_{\rm
[s]}$, 25\% Fe$_{\rm [s]}$, and 10\% Mg$_2$SiO$_4$$_{\rm [s]}$ with Al and Ti -- oxide impurities.  The chemical composition of
the dust is also imprinted in the element abundances of the gas phase and the molecular abundances adjust accordingly.

\subsection{Dust clouds growing from small $\longrightarrow$ lage}

\begin{figure*}
\label{fig:2D}
\begin{center}
\begin{tabular}{cc}
{\bf a)} \hspace*{0.5cm} $t=0.369$\,s & {\bf b)} \hspace*{0.5cm} $t=0.391$\,s \\
\epsfig{file=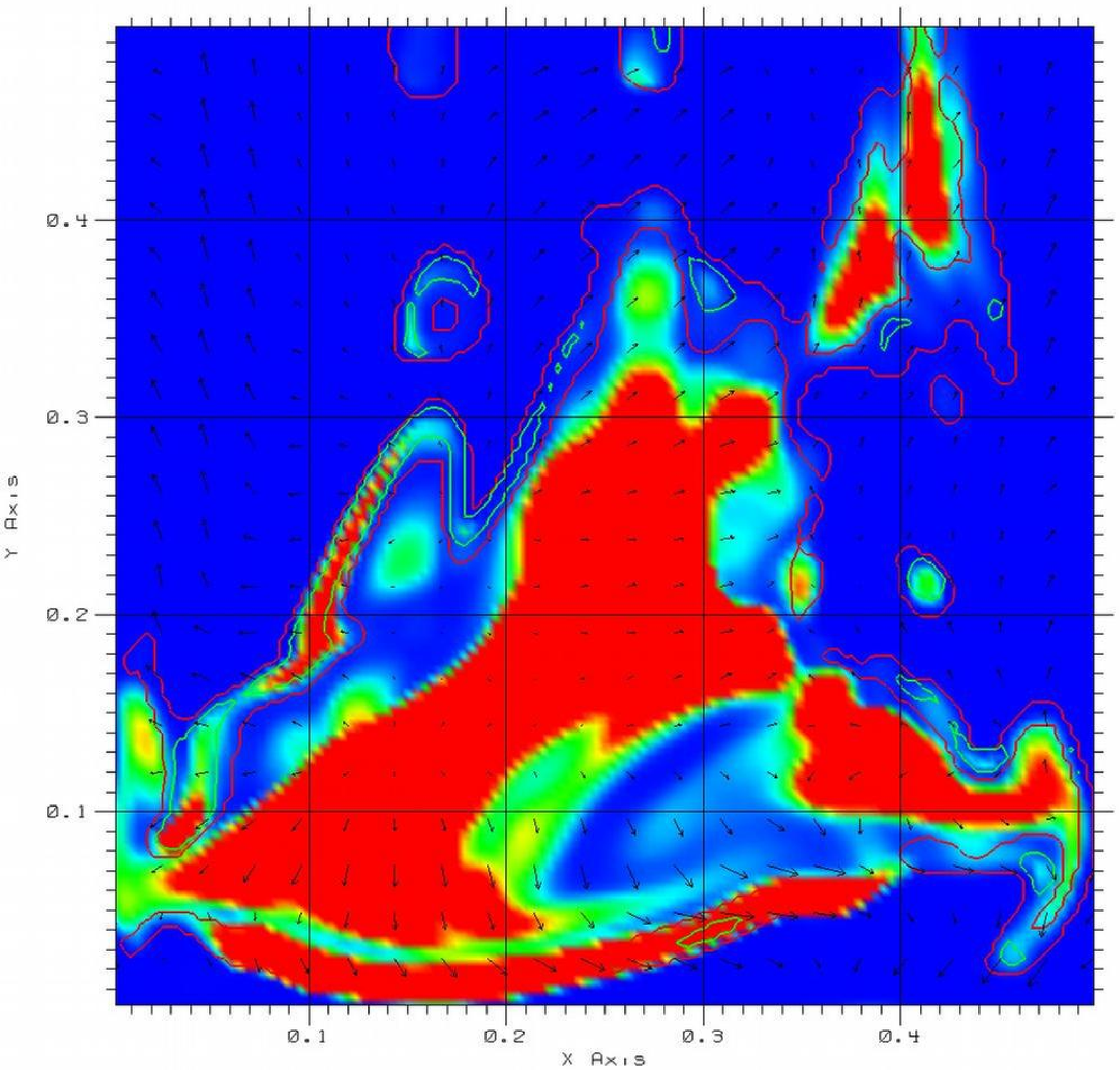, scale=0.22} &
\epsfig{file=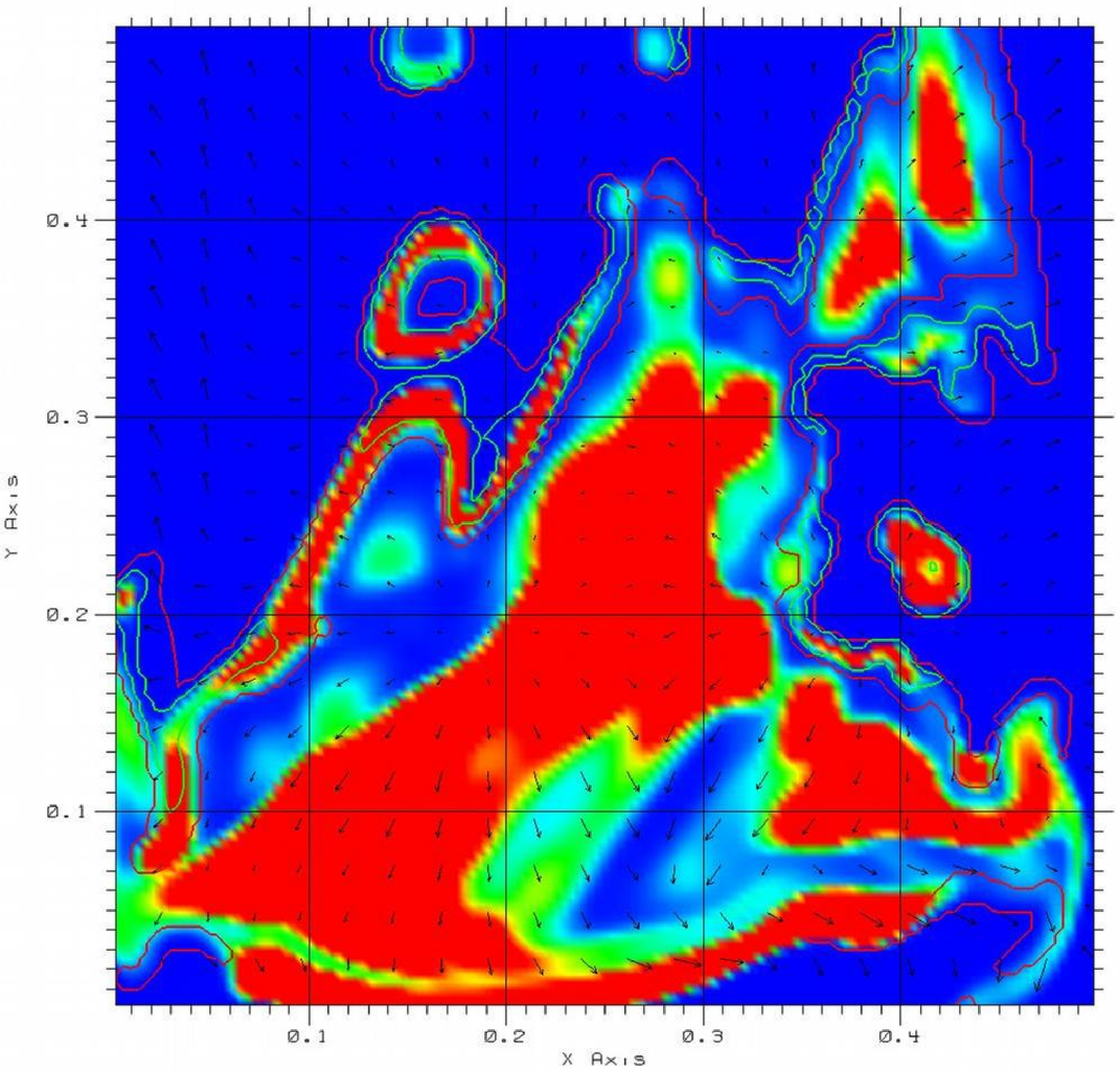, scale=0.22} \\
{\bf c)} \hspace*{0.5cm} $t=6.4$\,s & {\bf d)} \hspace*{0.5cm} $t=25$\,s\\
\epsfig{file=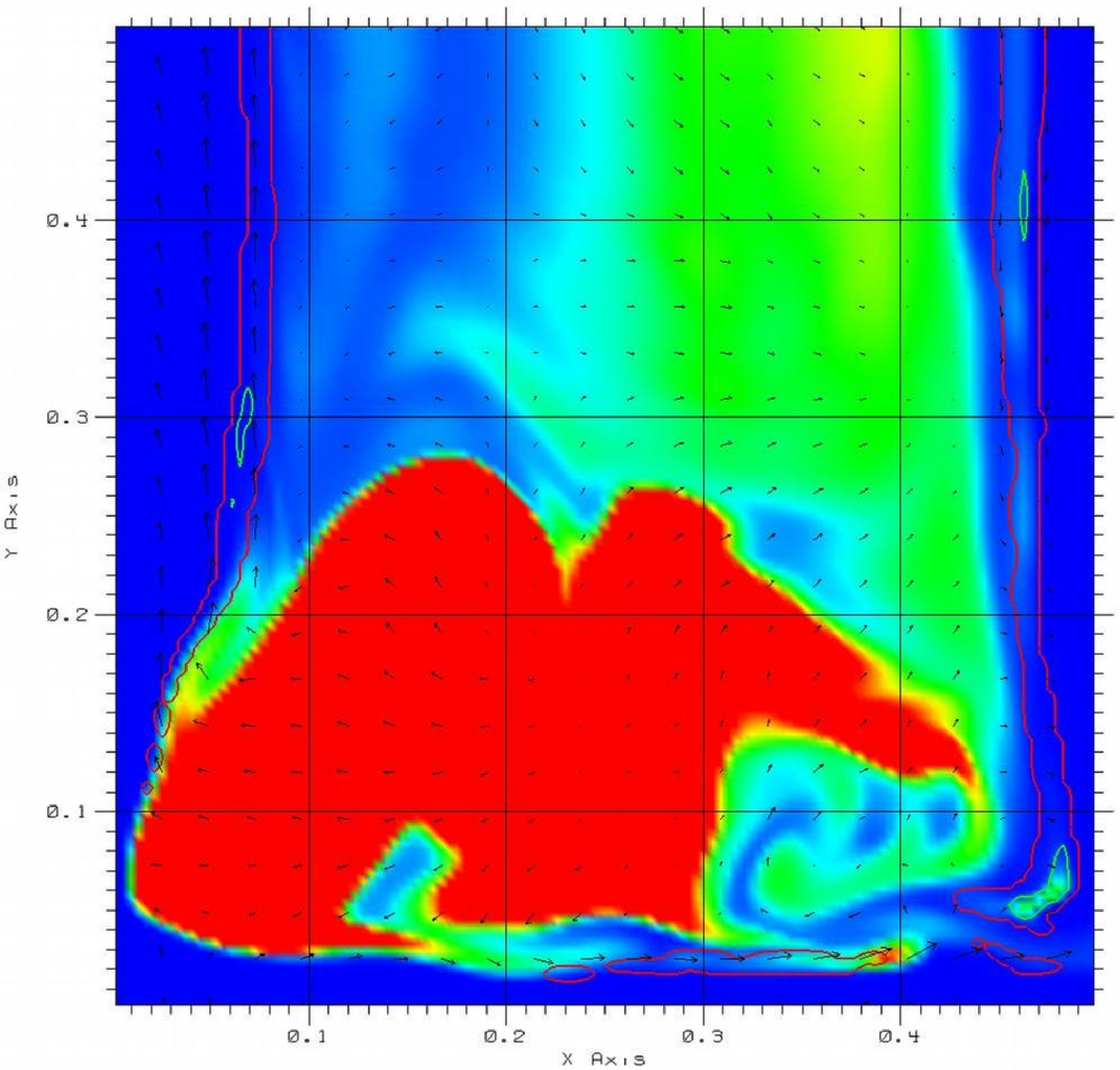, scale=0.22} &
\epsfig{file=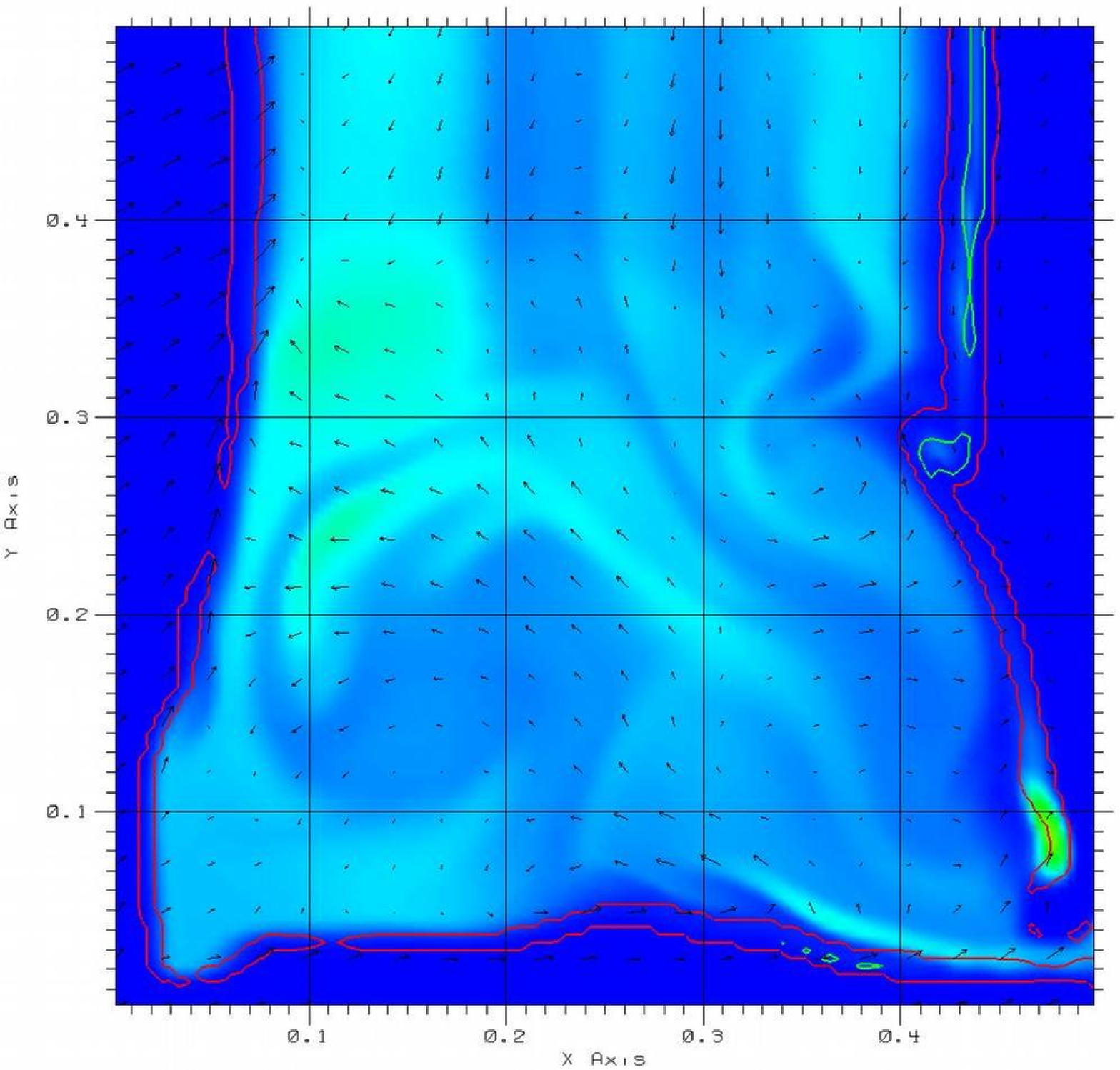, scale=0.22}\\
\end{tabular}
\caption{$T_{\rm ref}=2100$K, $T_{\rm RE}=1980$K, $g = 10^5$ g/cm$^2$,
$\rho_{\rm ref}=3.16\cdot 10^{-4}$g/cm$^3$, M=1 ($u_{\rm ref}=3\cdot
10^5$cm/s), $t_{\rm ref}=0.3$, $l_{\rm ref}=10^5$cm [numerical parameter:
$N_x\times N_y= 128 \times 128$, $N_k=500$, $\Delta x = 3.94\cdot
10^2$cm, $l_{\rm max}= l_{\rm ref}/2$]\newline $\log n_{\rm d}$ number
of dust particles -- false color (red = max, blue = min), $\log J_*$
nucleation rate -- contour lines (green = max, red = min), $\vec{v} =
v_x + v_y$ fluid velocity -- vector arrows.}
\end{center}
\end{figure*}

1D simulations allow a very detailed investigation of the physio-chemical interactions but fail to explain 2-- or 3--dimensional
phenomena like cloud or vortex formation as it was suggested by observations of Saturn's and Jupiter's atmospheres. The study of the
formation and the possible appearance of large scale dust clouds in substellar atmospheres was performed utilizing a model for
driven turbulence.  Turbulence is modeled by the superposition of $N_k$ modes each having a Kolmogoroff velocity amplitude (for
details Helling et al. 2004). Convection is believed to drive the turbulence in a real substellar atmosphere.  The 2D simulation
(Fig.~\ref{fig:2D}) is started from a homogeneous, dust free medium which constantly is disturbed by the turbulence driving from
the left, the right and the bottom side during the simulation.

During the initial phase of the simulations small scale nucleation events occur where turbulence causes the local temperature to
drop below the nucleation threshold (compare Sect.~\ref{sec:td}). Observe that a locally maximum $J_*$ in Fig.~\ref{fig:2D} a) is
immediately followed by an subsequent increase in $n_{\rm d}$ (panel b). These dusty areas tend to increase as the fluid motion
transports the dust into areas with a still undepleted gas phase. And indeed, as the simulation proceeds in time, larger and more
compact cloud-like dust structures are formed (panel c). These large-scale structures are the result of the hydrodynamic fluid
motion which gathers more and more dust also by the vortices appearing in the velocity field. Note the mushroom-like structure
evolving e.g. in the right lower corner in panel c). Strong radiative cooling causes these dusty areas to become
isothermal. Eventually, the cloud will leave the test volume or it will get disrupted and the small-scale fragments move out of
sight. The whole dust cloud formation cycle can start again only if metal species (Mg, Si, Fe, Al, Ti, O) are replenished from
outside like it has to be expected to occur by convection in a substellar atmosphere (panel d).

\section{Conclusion} Substellar atmospheres, i.e. atmospheres of giant gas planets and brown dwarfs, possess two scale regimes:
(i) the large-scale convection causing element replenishment of the upper atmosphere and the counteracting gravitational settling of dust
causing an element depletion of the upper atmosphere\\ (ii) small-scale turbulence and dust formation establishing a feedback
loop.

Regime (i) can be considered as quasi-static with view on the dust formation process. The dynamic regime (ii) is determined by the
turbulence initiating spot-like dust formation on small spatial scales. The strong radiative cooling by dust results in an
implosion of the dust forming areas which eventually become isothermal. Velocity disturbances occur which feed back into the
turbulent fluid field. The resulting mesoscopic flow gathers the dust in even larger, more compact cloud-like structures. The
final chemical composition of the grain is a function of the grains history.

\subsection*{Acknowledgements} The ESA research fellowship program at ESTEC is acknowledged. W. Traub and M. Fridlund are thanked
  for discussion on aerosols. Most of the literature search has been performed with the ADS system.

\end{document}